\documentstyle[12pt,supercite]{article}
\def\abstracts#1#2#3{{
	\centering{\begin{minipage}{4.62in}\baselineskip=13pt
        \small
        \centerline{\bf Abstract}
        \vspace*{0.2cm}                
        \parindent=0pt #1\par
        \parindent=18pt #2\par
        \parindent=15pt #3
        \end{minipage} }\par}}
\begin{document}
\vspace*{-1.5cm}
\hfill Mainz preprint KOMA-96-32\\
\mbox{}\hfill          July 1996\\
\vspace*{1.5cm}
\centerline{\LARGE \bf Three-dimensional 3-state Potts model}\\[0.3cm]
\centerline{\LARGE \bf revisited with new techniques}\\[0.4cm]
\vspace*{0.2cm}
\centerline{\large {\em Wolfhard Janke$^1$ and Ramon Villanova$^2$\/}}\\[0.4cm]
\centerline{\large {\small $^1$ Institut f\"ur Physik,
Johannes Gutenberg-Universit\"at Mainz,}}
\centerline{    {\small Staudinger Weg 7, 55099 Mainz, Germany }}\\[0.5cm]
\centerline{\large {\small $^2$ Matem\`atiques Aplicades,
                    Universitat Pompeu Fabra,}}
\centerline{    {\small La Rambla 32, 08002 Barcelona, Spain }}\\[0.5cm]
\abstracts{}{
We report a fairly detailed finite-size scaling analysis of the first-order
phase transition in the three-dimensional 3-state Potts model on cubic 
lattices with emphasis on recently introduced quantities whose infinite-volume 
extrapolations are governed {\em only} by exponentially small terms. In these
quantities no asymptotic power series in the inverse volume are involved 
which complicate the finite-size scaling behaviour of standard observables
related to the specific-heat maxima or Binder-parameter minima. Introduced 
initially for strong first-order phase transitions in $q$-state Potts models
with ``large enough'' $q$, the new techniques prove to be surprisingly 
accurate for a $q$ value as small as 3. On the basis of the high-precision
Monte Carlo data of Alves {\em et al.} [Phys. Rev. {\bf B43} (1991) 5846],
this leads to a refined estimate of $\beta_t = 0.550\,565(10)$ for the 
infinite-volume transition point. 
}{}
\vspace*{1.5cm}
\noindent PACS numbers: 05.50.+q, 75.10.Hk, 64.60.Cn\\[0.5cm]
e-mail: ~janke@miro.physik.uni-mainz.de\\
WWW: http://www.cond-mat.physik.uni-mainz.de/\~~$\!\!$janke
\thispagestyle{empty}
\newpage
\pagenumbering{arabic}
%
\section{Introduction}
%
The three-dimensional (3D) 3-state Potts ferromagnet serves as an important
model in both condensed matter as well as high-energy physics \cite{wu}. 
Experimental realizations are structural phase transitions in some crystals,
and theoretically this model has attracted much interest as a simple effective
model of finite-temperature pure-gauge QCD. Consequently it has been studied
in the past few years by many authors using quite a variety of different 
techniques \cite{straley,kj75,mbe79,bs79,herrmann79,kjm79,oi82,wv87,gkp89,%
fmou90,abv91,ya93,fuok89,huma89,blm90,patkos,bonfim91,leko91,detar93,vol93,%
schmidt,vinti}. By analyzing Monte Carlo (MC) simulations with the help of 
standard finite-size scaling (FSS) methods \cite{fss_1,borgs,fss_2,fss_3}
the characteristic parameters of the phase transition (transition temperature,
latent heat, etc.\ ) have been estimated with varying accuracy. As a result 
there is by now general consensus that this model undergoes on a simple cubic
lattice a weak first-order phase transition from a three-fold degenerate 
ordered low-temperature phase to a disordered phase at high temperatures. 
 
Most high-precision MC results are based on the FSS of the specific-heat
maxima, the Binder-parameter minima, or the partition function zeros. 
For some of these observables the pseudo-transition points, $\beta_t(L)$, 
exhibit a non-monotonic FSS behaviour with a peculiar dip around linear 
lattice sizes $L=30$ for, e.g., the specific-heat data. This clearly indicates
that for the standard observables the asymptotic FSS region is not reached 
until $L > 30$. For large systems with periodic boundary conditions, as a 
consequence of phase coexistence at a first-order phase transition, the FSS
behaviour of these observables is governed by asymptotic series expansions in
inverse powers of the volume $V = L^D$ \cite{fss_1,borgs,fss_2,fss_3}. In 
addition there are also further corrections that decrease exponentially with
the system size \cite{bj92,jan93}. They originate from finite-size effects 
$\propto \exp(-L/L_0)$ in the pure phases, where $L_0$ is of the order of the
finite (pure phase) correlation lengths $\xi_{o,d}$ in the ordered and
disordered phase, and from contributions $\propto \exp(-2\sigma_{od} L^{D-1})$ 
of the two-phase region, where $\sigma_{od}$ is the (reduced) interface 
tension between the two phases. For 3D systems the latter correction 
decays asymptotically much faster and, in general, both types of exponential
corrections are always weaker than the power-law terms for large $L$.
For relatively small system sizes with $L$ of the order of $\xi_{o,d}$ or 
$\sigma_{od}^{-1/(D-1)}$, however, it is a priori not clear which type of 
scaling behaviour will dominate. A possible explanation for the observed dip
in the scaling behaviour of the 3D 3-state Potts model is therefore that for 
$L < 30$ the exponential terms dominate and then rapidly die out, giving way
to an inverse volume dependence for $L > 30$.

Consequently, in order to extract the infinite-volume transition point,
$\beta_t$, from a pure power-law ansatz in the inverse volume, 
in previous studies only the data from lattices with $L \ge 30$ have been 
taken into account. Of course, this is an expensive task because the computer
time to generate accurate enough data points for such lattice sizes becomes
exceedingly large, even with refined MC techniques for first-order phase 
transitions such as multicanonical \cite{bn92} or multibondic \cite{jk95}
algorithms which reduce autocorrelation times significantly. 

Recently a new set of observables was proposed which do not show any 
power-law terms in their FSS behaviour \cite{bj92,jan93}. It is thus 
governed {\em only\/} by exponentially small terms. For lattices with 
periodic boundary conditions and large enough $q$, this could be proven 
exactly \cite{borgs}. In Refs.~\cite{bj92,jan93} the new set of observables
was first used to study the first-order phase transitions of the 
two-dimensional $q$-state Potts model with $q=5$, 8, and 10. It was found 
that the new methods yield surprisingly accurate results already for very 
small lattice sizes and even for the extremely weak first-order transition
of the 2D 5-state model. Since this might be a fortuitous accident of the 
2D model we found it worthwhile to explore the accuracy of the new methods 
in the 3D case as well. The purpose of this paper is thus to test the validity
of the new observables for three-dimensional lattices and for a $q$ value as
small as 3. We shall see that, due to the absence of power-law corrections, 
the dip in the new pseudo-transition points $\beta_t(L)$ indeed disappears, 
and a good fit to extract the infinite-volume transition point 
$\beta_t \equiv \beta_t(\infty)$ becomes feasible without the need of 
extremely large lattices. 

The rest of the paper is organized as follows. In Sec.~2 we recall the 
definition of the model and briefly summarize previous results, and in Sec.~3
we recapitulate the definition of the new observables. The results of our
analysis are presented in Sec.~4, and in Sec.~5 we conclude with a brief
discussion.
%
\section{Model and previous results}
%
We use the standard definition of the $3$-state Potts model \cite{wu},
\begin{equation}
   Z = \sum_{\{\sigma_i\}} e^{-\beta E} = 
                  \sum_{E} N(E) e^{-\beta E}, \; \; \; \;
   E = -\sum_{\langle ij \rangle} \delta_{\sigma_i, \sigma_j},
   \; \; \; \sigma_i = 1,2,3,
\label{eq:zpotts}
\end{equation}
where $\beta = J/k_{\rm B}T$ is the inverse temperature in natural units, 
$N(E)$ is the number of configurations with energy $E$,
${\langle ij \rangle}$ indicates that the sum runs over nearest-neighbour
pairs, and $\delta_{\sigma_i, \sigma_j}$ is the Kronecker symbol. 
The energy per site is denoted by $e = E/V$, where $V = L^3$ is the volume of
the system. The data we use to test our observables was obtained in 
a previous MC simulation by Alves {\em et al.} (ABV) \cite{abv91}
who employed heat-bath updating for lattices of size $L=10$, 12, 14, 18, 22,
24, 30, and 36, with periodic boundary conditions. For all the lattice sizes,
the simulations were performed at $\beta_{\rm MC}=0.55059$, with 
$50 \times 10^6$ energy measurements for $L=10$, 12, and 14, and 
$20 \times 10^6$ measurements for the larger lattice sizes. 

By means of standard reweighting procedures \cite{rew}, 
the data was used to compute the maximum $C_{\rm max} = C(\beta_{C_{\rm max}})$
of the specific heat, 
$C(\beta)=\beta^2 V(\langle e^2 \rangle - \langle e \rangle^2)$, and the 
Fisher zeros of the partition function. Finite-size scaling was finally 
applied to both the specific-heat maxima and the partition function zeros
in order to extract the transition point and the FSS exponents.  
Due to the above described problem with the non-monotonic scaling behaviour
of $\beta_{C_{\rm max}}(L)$, ABV combined their data with those of 
Ref.~\cite{fmou90} on larger lattices and extracted their final estimate%
\footnote{The notation of ABV follows the high-energy physics conventions. The 
translation into the present notation, usually used in condensed matter 
physics, reads $\beta_t = (3/2) \beta_t^{\rm (ABV)}$. Similarly the latent
heats are related by $\Delta e = 2 \Delta e^{\rm (ABV)} \equiv 2 \ell$.} of 
$$\beta_t = 0.550\,523 \pm 0.000\,010 \mbox{~~~~(ABV)}$$ 
from a power-law fit, $\beta_{C_{\rm max}} = \beta_t + c/V$, using only the 
largest lattices of size $L = 30$, 36, 42, and 48. The latent heat,
$$\Delta e \equiv e_d - e_o = 0.160\,62 \pm 0.000\,52 \mbox{~~~~(ABV)},$$
was obtained from the FSS of the specific-heat maxima on smaller lattices of 
size $L = 22$, 24, 30, and 36, since the scaling behaviour of $C_{\rm max}$ 
turned out to be better behaved than that of the maxima locations. An overview
of other previous estimates of $\beta_t$ is given in 
Table~\ref{tab:previous_bt}.
\begin{table}

\caption[a]{{\em The transition point $\beta_t$ of the 3-state Potts model
 on the simple cubic lattice in units $k_{\rm B} = 1 = J$}}

\begin{center}

\begin{tabular}{ll}
\hline
\multicolumn{1}{c}{$\beta_{\rm t}$} & \multicolumn{1}{c}{author(s)} \\
\hline
0.534\,8       & Straley (1974) \cite{straley}     \\            
0.559\,6(54)   & Kim and Joseph (1975) \cite{kj75} \\            
0.547\,49(52)  & Miyashita {\em et al.\/} (1979) \cite{mbe79}\\  
0.550          & Bl\"ote and Swendsen (1979) \cite{bs79} \\      
0.550\,06(91)  & Herrmann (1979) \cite{herrmann79} \\            
0.550\,40(19)  & Knak-Jensen and Mouritsen (1979) \cite{kjm79}\\ 
0.552          & Ono and Ito (1982) \cite{oi82} \\               
0.550\,59(2)   & Wilson and Vause (1987) \cite{wv87} \\          
0.550\,62(3)   & Gavai {\em  et al.\/} (1989) \cite{gkp89} \\    
0.550\,545(45) & Fukugita {\em  et al.\/} (1990) \cite{fmou90}\\ 
0.550\,523(10) & Alves {\em et al.\/} (1991) \cite{abv91} \\     
0.550\,479(61) & Yamagata (1993) \cite{ya93} \\                  
0.550\,565(10) & this work \\
\hline
\end{tabular}

\end{center}

\label{tab:previous_bt}

\end{table}
%
\section{The new observables}
\label{sec:newobser}
%
The definition of the new observables is based on the observation made in 
Ref.~\cite{borgs} that for a $q$-state Potts model on a lattice with 
periodic boundary conditions, provided  $q$ is large enough, the partition
function can be written as
\begin{equation}
 Z_{\rm per}(V,\beta)= \left[ \sum_{m=0}^q e^{-\beta f_m(\beta)V} \right]
 [1+O(V e^{-L/L_0})],
 \label{eq:zper}
\end{equation}
where $L_0 < \infty$ is a constant, $L$ is the linear length of the lattice,
and $f_m(\beta)$ is the metastable free-energy density of the phase $m$, where
$m=0$ is associated with the disordered and $m=1,\dots,q$ with the $q$
ordered phases. It can be defined in such a way that it is equal to the 
idealized infinite-volume free-energy density $f(\beta)$ if $m$ is stable and
strictly larger than $f(\beta)$ if $m$ is unstable. At $\beta_t$ all $f_m$ 
are equal.

In Refs.~\cite{bj92,jan93} it is discussed how Eq.~(\ref{eq:zper}) leads 
to a definition of pseudo-transition points $\beta_{V/V'}$ for finite systems,
which deviate from the infinite-volume transition point $\beta_t$ only by 
exponentially small terms. These points are obtained by locating the maximum
of the {\em number-of-phases} observable
\begin{equation}
N(V,V',\beta) \equiv \left[ \frac{Z_{\rm per}(V,\beta)^\alpha}
                                 {Z_{\rm per}(V',\beta)}\right]^{1/(\alpha-1)},
\label{eq:N}
\end{equation}
where $\alpha = V'/V$. By inserting (\ref{eq:zper}) it is easy to see that 
$N(V,V',\beta)$ indeed counts the number of phases: $q$ at low temperatures, 
$q+1$ at phase coexistence, and $1$ in the disordered high-temperature phase.
Searching for the maximum of $N(V,V',\beta)$ as a function of $\beta$ amounts
to locating the crossing points of the internal energies per site for the two
lattices of different size,
\begin{equation}
     e(V,\beta_{V/V'}) = e(V',\beta_{V/V'}).
 \label{eq:betacross}
\end{equation}
As a consequence of (\ref{eq:zper}) it can be proven that, for large enough
$q$, the crossing points $\beta_{V/V'}$ are only exponentially shifted from
the infinite-volume transition point $\beta_t$. In contrast to the 
pseudo-transition points $\beta_{C_{\rm max}}(L)$, there are no powers of 
the inverse volume involved.

Also in Refs.~\cite{bj92,jan93}, yet another definition of pseudo-transition
points, $\beta_w(L)$, without power-law corrections was proposed which
are obtained by measuring the {\em ratio-of-weights},
\begin{equation}
 R_W(V,\beta) \equiv \frac{W_o}{W_d} \equiv \sum_{E<E_0} P_{\beta}(E) / 
                     \sum_{E \geq E_0} P_{\beta}(E), 
\label{eq:R_W} 
\end{equation}
and solving the equation
\begin{equation}
 R_W(V,\beta)|_{\beta=\beta_w} = q = 3.
\label{eq:ratiowe} 
\end{equation}
Here, $P_{\beta}(E)$ stands for the probability distribution
$P_{\beta}(E) = N(E) e^{-\beta E}/Z$, with $Z$ and $N(E)$ defined in 
Eq.~(\ref{eq:zpotts}). The parameter $E_0$ in Eq.~(\ref{eq:R_W}) is defined 
by reweighting the probability distribution to the point $\beta_P$ where the 
two peaks of $P_{\beta}(E)$ have equal height and then taking $E_0$ as the 
energy where $P_{\beta_P}(E)$ has the minimum between the two peaks. This 
defines still another sequence of pseudo-transition points $\beta_P$ which,
however, is {\em not} expected to exhibit a FSS behaviour free of $1/V$ powers.
%
\section{Results}
%
\subsection{Number-of-phases criterion}
 From Eq.~(\ref{eq:betacross}) it is clear that the numerical
determination  of $\beta_{V/V'}$ requires handling simultaneously
two different lattices sizes. Once they are chosen, searching for 
the minimum of 
\begin{equation}
     |e(V,\beta_{V/V'}) - e(V',\beta_{V/V'})| = 0,
\end{equation}
using standard numerical minimizing routines combined with reweighting 
procedures, leads to the $\beta_{V/V'}$ collected in Table~\ref{tab:betas_V}
for all possible pairs of sizes $L$ and $L'$. Here and in the following 
statistical error bars are estimated by means of the jack-knife 
method \cite{jack} on the basis of 20 blocks.
%
%
%
\scriptsize
\begin{table}[tb]
\centering
\caption[{\em Nothing.}]
 {{\em  Pseudo-transition points $\beta_{V/V'}$  
        for pairs of lattices of size $L$ and $L'$}}
\vspace{3ex}
\begin{tabular}{l|llllll}
\multicolumn{1}{l|}{$\!L \setminus L'\!\!$}  &
\multicolumn{1}{c}{14} &
\multicolumn{1}{c}{18} &
\multicolumn{1}{c}{22} &
\multicolumn{1}{c}{24} &
\multicolumn{1}{c}{30} &
\multicolumn{1}{c}{36} \\ \hline 
 10 & 0.552309(24) & 0.551757(15) & 0.551421(15) &
             0.5513025(75) & 0.5510784(59) & 0.5509544(50) \\ \\
 12 & 0.551798(41) & 0.551405(21) & 0.551142(13) & 
             0.5510459(72) & 0.5508701(48) & 0.5507715(56) \\ \\
 14 &              & 0.551220(30) & 0.550998(17) &  
             0.5509160(95) & 0.5507778(53) & 0.5507039(57) \\ \\
 18 &              &              & 0.550808(41) &
             0.550751(18)  & 0.5506733(90) & 0.5506343(63) \\ \\
 22 &              &              &              & 
             0.550649(66)  & 0.550624(15)  & 0.5506043(87) \\ \\
 24 &              &              &              &
                           & 0.550618(14)  &  0.5505995(83)\\ \\
 30 &              &              &              &
                           &               &  0.550586(18) \\ \\
\end{tabular}
\label{tab:betas_V}
\end{table}
\normalsize

\begin{figure}[bhp]
\vskip 7.0truecm
\includegraphics{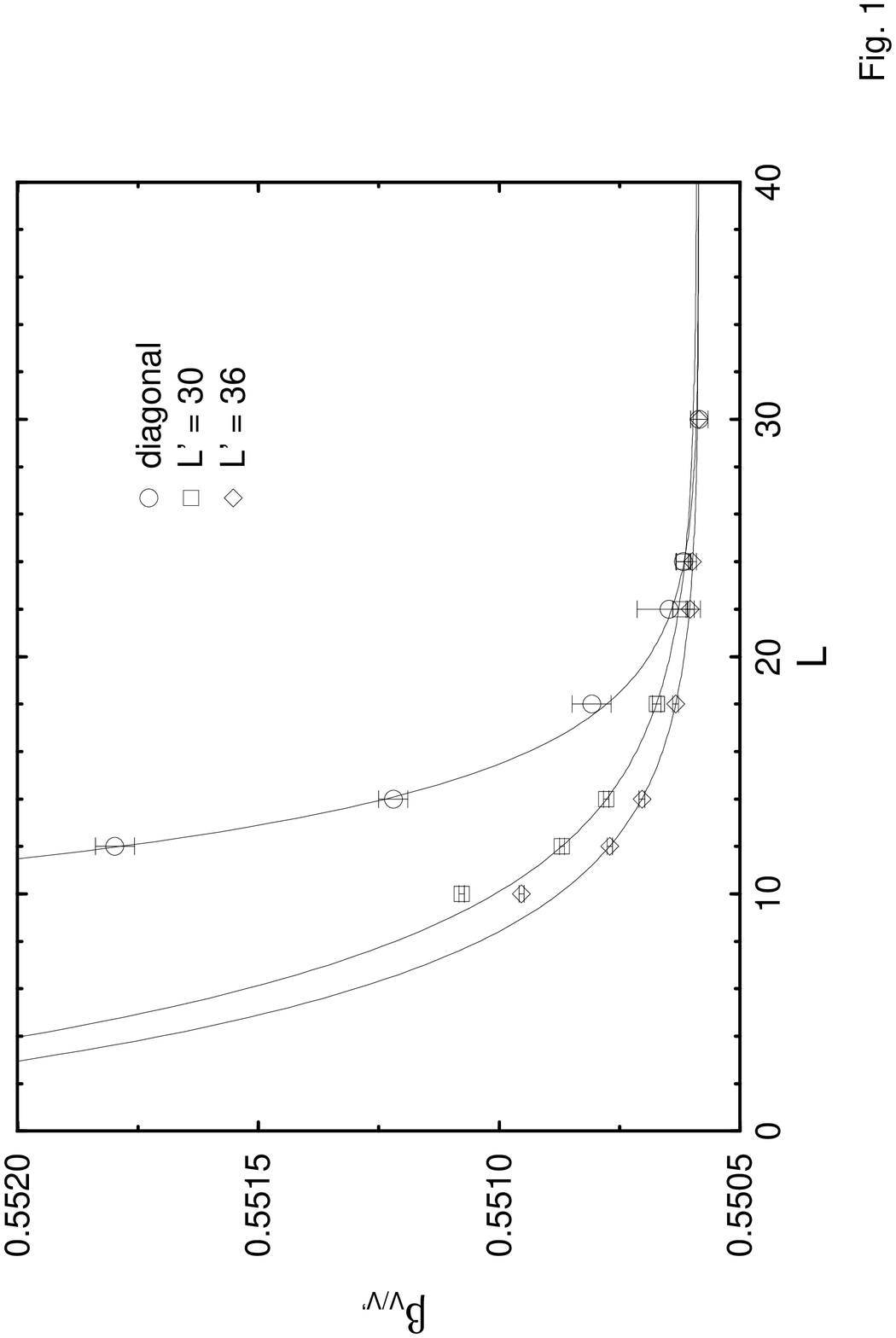}
\caption[a]{{\em Finite-size scaling of the pseudo-transition points
$\beta_{V/V'}$ of the number-of-phases criterion together with exponential
fits of the form $\beta_{V/V'} = \beta_t + a \exp(-bL)$.}}
\label{fig:beta_vv}
\end{figure}

The question arises on how to extract an infinite-volume transition point
from the results given in Table~\ref{tab:betas_V}. Considering that, as 
discussed in Sec.~\ref{sec:newobser}, the pseudo-transition points 
$\beta_{V/V'}$ are left with only exponential deviations from $\beta_t$, 
we first tried a fit of the form
\begin{equation}
      \beta_{V/V'}(L) = \beta_t + a_1 e^{-b_1 L},
\label{eq:fit1}
\end{equation}
with $L' > L$ kept fixed. It seems evident that the bigger $V$ and $V'$ the
better the fit. By keeping $L'=36$ fixed and using $L = 12,\dots,30$ 
(i.e., the last column in Table~\ref{tab:betas_V} without the $L=10$ 
entry) we obtain an estimate of 
\begin{equation}
\beta_t = 0.550\,586 \pm 0.000\,010 \mbox{~~~~(number-of-phases criterion)},
\label{eq:bt1} 
\end{equation}
with $a_1 = 0.0027(10)$, $b_1=0.224(34)$, and a goodness-of-fit parameter 
of $Q=0.99$ (corresponding to a total chi-squared of $\chi^2 = 0.080$ with
3 degrees of freedom (dof)). The goodness of the fit can be visually inspected
in Fig.~\ref{fig:beta_vv}. Strictly speaking the parameters in (\ref{eq:fit1})
should still depend on the (larger) lattice size $L'$. To estimate this
effect we performed a similar fit with $L'=30$ fixed, using the crossing 
points with the smaller lattices of size $L=12,\dots,24$. As a result we
obtained a consistent estimate of $\beta_t = 0.550\,590(20)$, 
$a_1 = 0.00312(89)$, $b_1=0.201(28)$, and $Q=0.95$. 

As far as the systematic
FSS corrections are concerned it would be more reasonable to extrapolate the
$\beta_{V/V'}$ along the diagonal of Table~\ref{tab:betas_V}, where 
$L'/L \approx 1.1 \dots 1.3$, such that both, $L$ {\em and} $L'$, are sent
to infinity simultaneously. From the point of statistical errors, however,
this procedure is somewhat problematic. First, the crossing points of
energies for only slightly different lattice sizes have naturally the
biggest errors since the slopes $d e(V,\beta) /d\beta$ and 
$d e(V',\beta) /d\beta$ differ only little. This is clearly reflected in
Table~\ref{tab:betas_V}. Second, and more difficult to take properly into
account in principle, the data for, e.g., $\beta_{12^3/14^3}$ and 
$\beta_{14^3/18^3}$ are correlated since both involve the energy on the 
$14^3$ lattice. To avoid this correlation one would have to simulate many 
more different lattice sizes. We nevertheless also tried this type of fit 
and, using all 6 data points along the diagonal (with $L$ in (\ref{eq:fit1})
being the smaller of the two lattice sizes), obtained 
$\beta_t = 0.550\,586(15)$, $a_1 = 0.048(16)$, and $b_1=0.307(26)$, with 
$Q=0.79$. It is gratifying that the extrapolated values of $\beta_t$ are 
in perfect agreement for all three sequences of pseudo-transition points, 
even though the approach to the infinite-volume limit shown in 
Fig.~\ref{fig:beta_vv} looks quite different, in particular for the last 
sequence along the diagonal.

\subsection{Ratio-of-weights criterion}
\begin{figure}[bhp]
\vskip 7.0truecm
\includegraphics{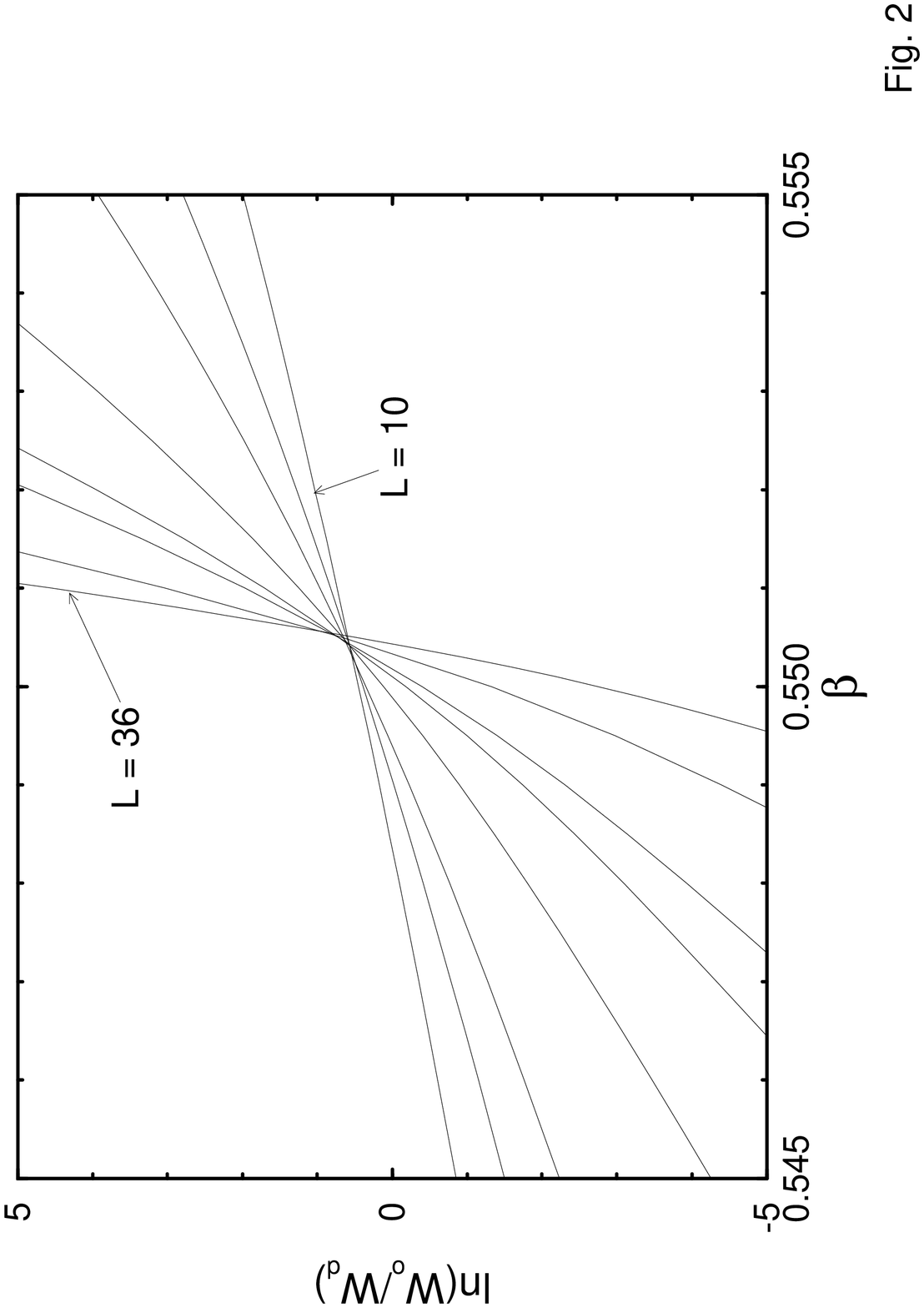}
\caption[a]{{\em The logarithm of the weight ratio as a function of the 
inverse temperature for lattices of size $L=$ 10, 12, 14, 18, 22, 24, 30, and
36.}}
\label{fig:w_ratio}
\end{figure}

 From a technical point of view, the second new definition (\ref{eq:ratiowe}) 
of pseudo-transition points, $\beta_w$, is probably somewhat easier to
evaluate numerically since it involves only one lattice size at a time.
A slight complication arises, however, from the fact that one has first 
to estimate $E_0$, the energy cut separating the ordered from the 
disordered phase. In order to determine $E_0$, we proceeded as follows. 
For a given lattice size $L$, we reweight the probability distribution of the
energy to the point $\beta_P$ where the ordered and disordered peaks are of 
equal height,
\begin{equation}
 P_{1,{\rm max}} = P_{\beta_P}(E_{1,{\rm max}})
                 = P_{\beta_P}(E_{2,{\rm max}}) = P_{2,{\rm max}}. 
\label{eq:beta_height}
\end{equation}
We then fix $\beta$ to $\beta_P$ and search for the minimum of 
$P_{\beta_P}(E)$ for $E$'s satisfying $E_{1,{\rm max}} < E < E_{2,{\rm max}}$.
Our $E_0$ is thus defined by $P_{\rm min} = P_{\beta_P}(E_0)$.
Once $E_0$ is determined it is kept fixed, and we reweight again over 
$\beta$ until Eq.~(\ref{eq:ratiowe}) is solved for $\beta=\beta_w$.
The behaviour of $R_W(V,\beta)$ as a function of $\beta$ is shown in
Fig.~\ref{fig:w_ratio}. 
%
%
%
\begin{table}[htbp]
\centering
\caption[{\em Nothing.}]
 {{\em  Table of $\beta_P$, $P_{\rm min}$ and the corresponding
        $2 \sigma_{od}$, together with $\beta_{w}$ }}
\vspace{3ex}
\begin{tabular}{r|llll}
\multicolumn{1}{c|}{$L$}  &
\multicolumn{1}{c}{$\beta_P$}  &
\multicolumn{1}{c}{$P_{\rm min}$}     &
\multicolumn{1}{c}{$2 \sigma_{od}$}   &
\multicolumn{1}{c}{$\beta_{w}$} \\
\hline
 10 & 0.550\,633(12)  & 0.8576(26) & 0.001\,536(30) &  0.552\,223(10)   \\[0.2cm]
 12 & 0.550\,638(20)  & 0.7911(34) & 0.001\,627(30) &  0.551\,6148(72)  \\[0.2cm]
 $14$ & 0.550\,585(11)  & 0.7310(43) & 0.001\,599(30) &  0.551\,2068(62)  \\[0.2cm]
 18 & 0.550\,572(15)  & 0.5869(81) & 0.001\,645(43) &  0.550\,878(10)   \\[0.2cm]
 22 & 0.550\,567(12)  & 0.4561(87) & 0.001\,622(39) &  0.550\,700(10)   \\[0.2cm]
 24 & 0.550\,552\,1(77) & 0.3984(94) & 0.001\,598(41) &  0.550\,658\,6(65)  \\[0.2cm]
 30 & 0.550\,556\,0(65) & 0.2291(54) & 0.001\,638(26) &  0.550\,596\,5(50)  \\[0.2cm]
 36 & 0.550\,566\,5(72) & 0.1155(66) & 0.001\,666(44) &  0.550\,581\,2(59)  \\[0.2cm]
\end{tabular}
\label{tab:beta_w}
\end{table}

The Table~\ref{tab:beta_w} collects the $\beta_P$ and $\beta_w$ introduced
in Eqs.~(\ref{eq:beta_height}) and (\ref{eq:ratiowe}) for $L=10,\dots,36$. 
Also given are the values of $P_{\rm min}$, employing the normalization
$P_{1,{\rm max}} = P_{2,{\rm max}} = 1$, as well as 
\begin{equation}
     2 \sigma_{od}(L) = -\ln(P_{\rm min})/L^2, 
\label{eq:sigma}
\end{equation}
which serves as a finite-volume estimator of the (reduced) interface tension
between the ordered and disordered phase \cite{binder} to be discussed below.

\begin{figure}[bhp]
\vskip 6.5truecm
\includegraphics{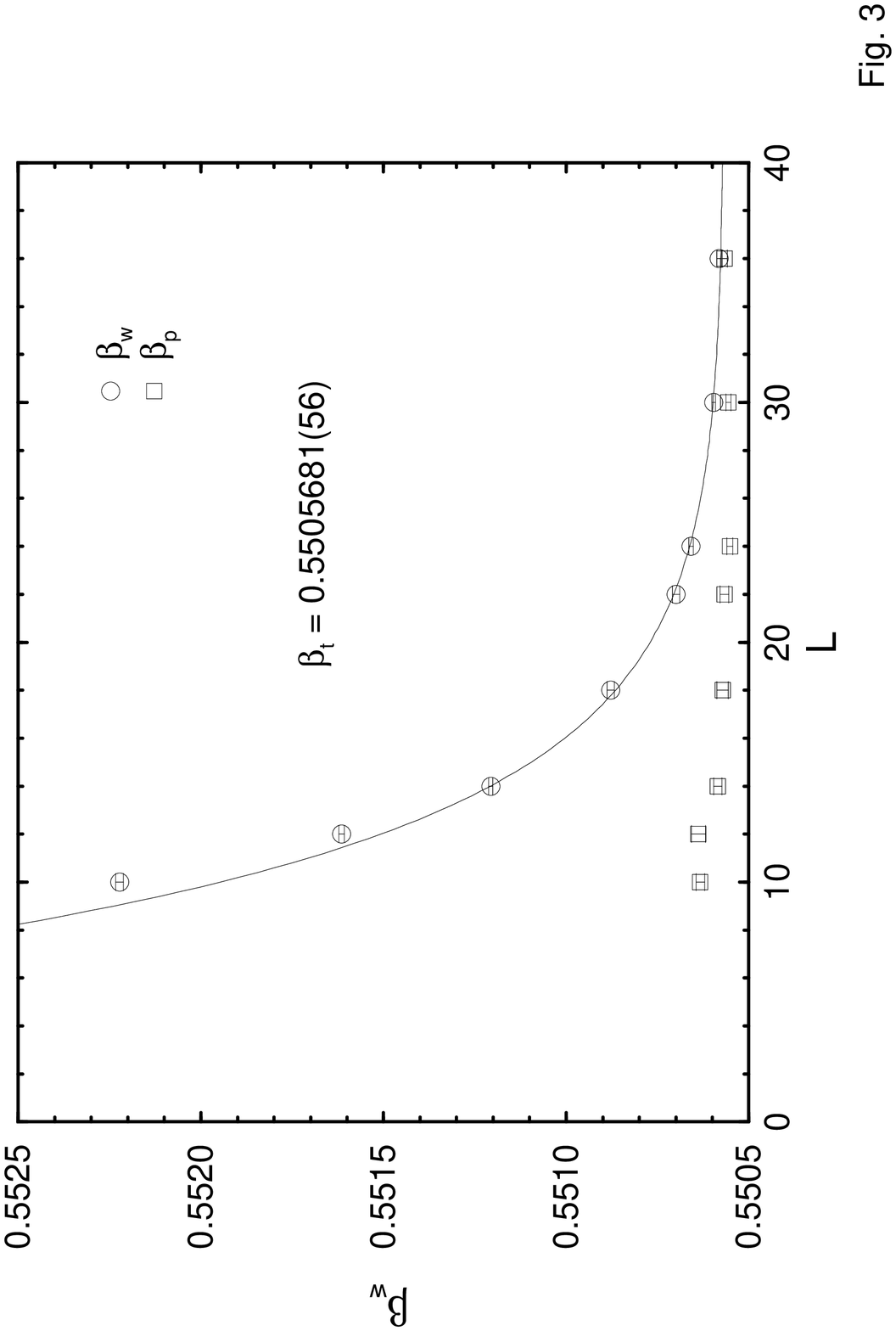}
\caption[a]{{\em Finite-size scaling of the pseudo-transition points
$\beta_{w}$ of the ratio-of-weights method together with an exponential
fit of the form $\beta_{w} = \beta_t + a \exp(-bL)$. Also shown are the 
points $\beta_p$ where the two peaks of the energy probability distribution 
are of equal height.}}
\label{fig:beta_w_2}
\end{figure}

In order to extract an estimate for the infinite-volume transition point,
we tried again an exponential fit of the form
\begin{equation}
     \beta_w(L) = \beta_t + a_2 e^{-b_2 L}.
\label{eq:fit2}
\end{equation}
Using the $\beta_w(L)$ for $L=14,\dots,36$ from Table~\ref{tab:beta_w},
we extract 
\begin{equation}
 \beta_t = 0.550\,568\,1 \pm 0.000\,005\,6 \mbox{~~~~(ratio-of-weights 
 criterion)},
\label{eq:bt2}
\end{equation}
with $a_2 = 0.00939(93)$, $b_2=0.1919(74)$, and $Q=0.41$. The data for
$\beta_w$ and $\beta_P$ together with the fit (\ref{eq:fit2}) are shown 
in Fig.~\ref{fig:beta_w_2}. Within error bars the estimates (\ref{eq:bt1})
and (\ref{eq:bt2}) are compatible with each other, but both are slightly 
higher than the previous estimate of ABV. Notice that also the estimates 
$b_1$ and $b_2$ are consistent, indicating that $L_0 = 1/b_{1,2}$ in 
(\ref{eq:zper}) is about $L_0 \approx 5$. This value is roughly one half of
the correlation length estimates $\xi_o \approx \xi_d \approx 10.2$, 
obtained recently in pure phase simulations of the 3-state Potts model on 
a $100^3$ lattice \cite{jk96}.

In general also the pseudo-transition points $\beta_P$ (where the two peaks
are of equal height) can be used to extract $\beta_t$. As mentioned at the
end of Sec.~3, the FSS behaviour should be qualitatively similar to 
$\beta_{C_{\rm max}}$, i.e., one expects an asymptotic power series in $1/V$.
A glance at Fig.~\ref{fig:beta_w_2} shows, however, that the $\beta_P$ for 
the three-dimensional 3-state Potts model are apparently almost constant. 
In fact, by simply taking the weighted average of the six values for 
$L=14,\dots,36$ (i.e., performing a trivial fit 
$\beta_P = \beta_t = {\rm const.}$), we obtain an estimate of 
\begin{equation}
\beta_t = 0.550\,562\,5 \pm 0.000\,003\,6 \mbox{~~~~(equal-peak-height 
locations)}, 
\label{eq:beqh}
\end{equation}
with $Q=0.16$, in good agreement with (\ref{eq:bt2}). If we omit the $L=14$ 
value, the average changes only little to $\beta_t = 0.550\,560\,0(38)$, 
with $Q=0.51$.

\begin{figure}[bhp]
\vskip 7.0truecm
\includegraphics{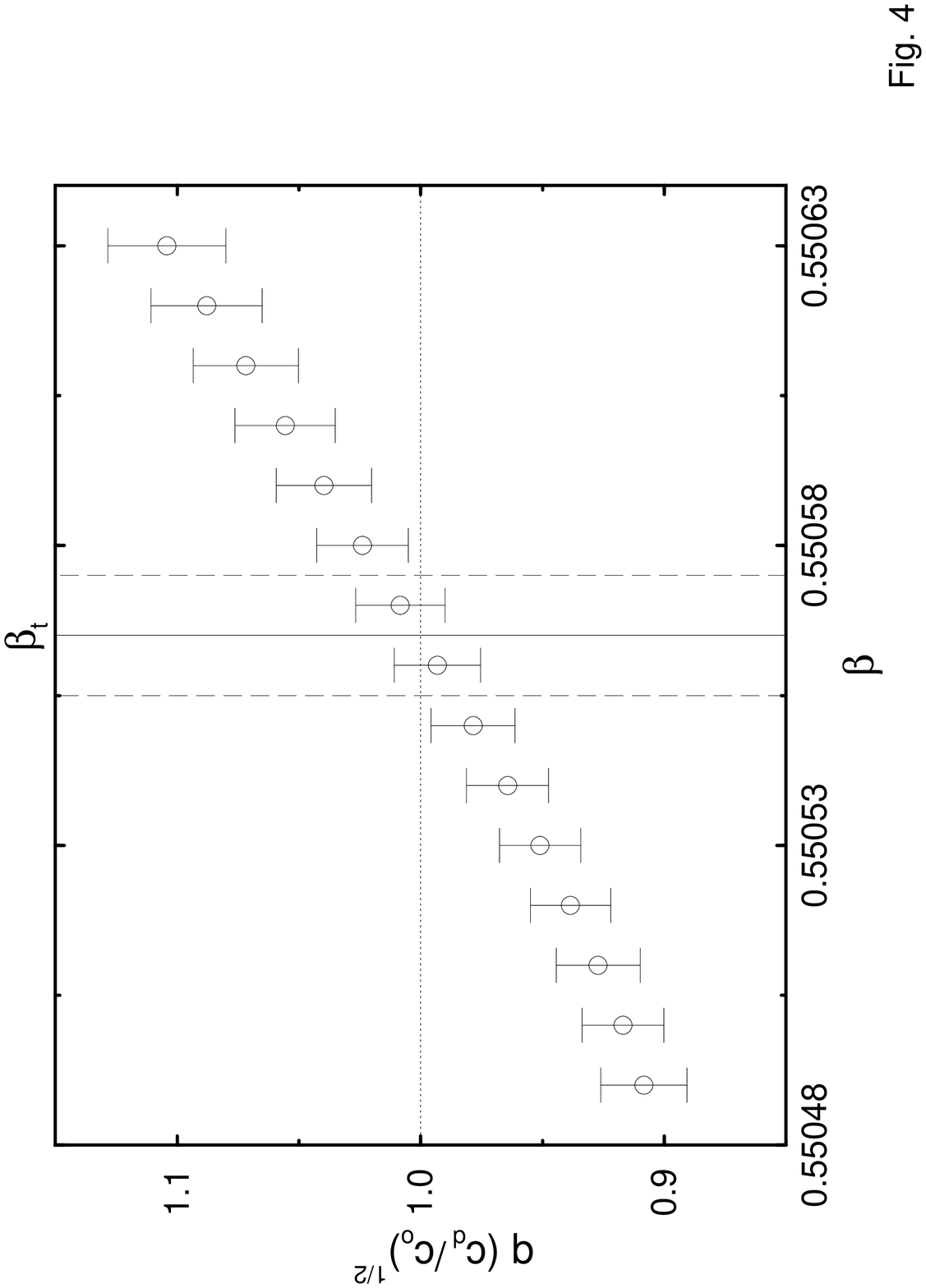}
\caption[a]{{\em Theoretical prediction (\ref{eq:R_H}) for the height ratio 
as a function of the inverse temperature. The vertical lines show the
infinite-volume transition point together with the statistical error bounds.}}
\label{fig:height_ratio}
\end{figure}

\begin{figure}[bhp]
\vskip 7.0truecm
\includegraphics{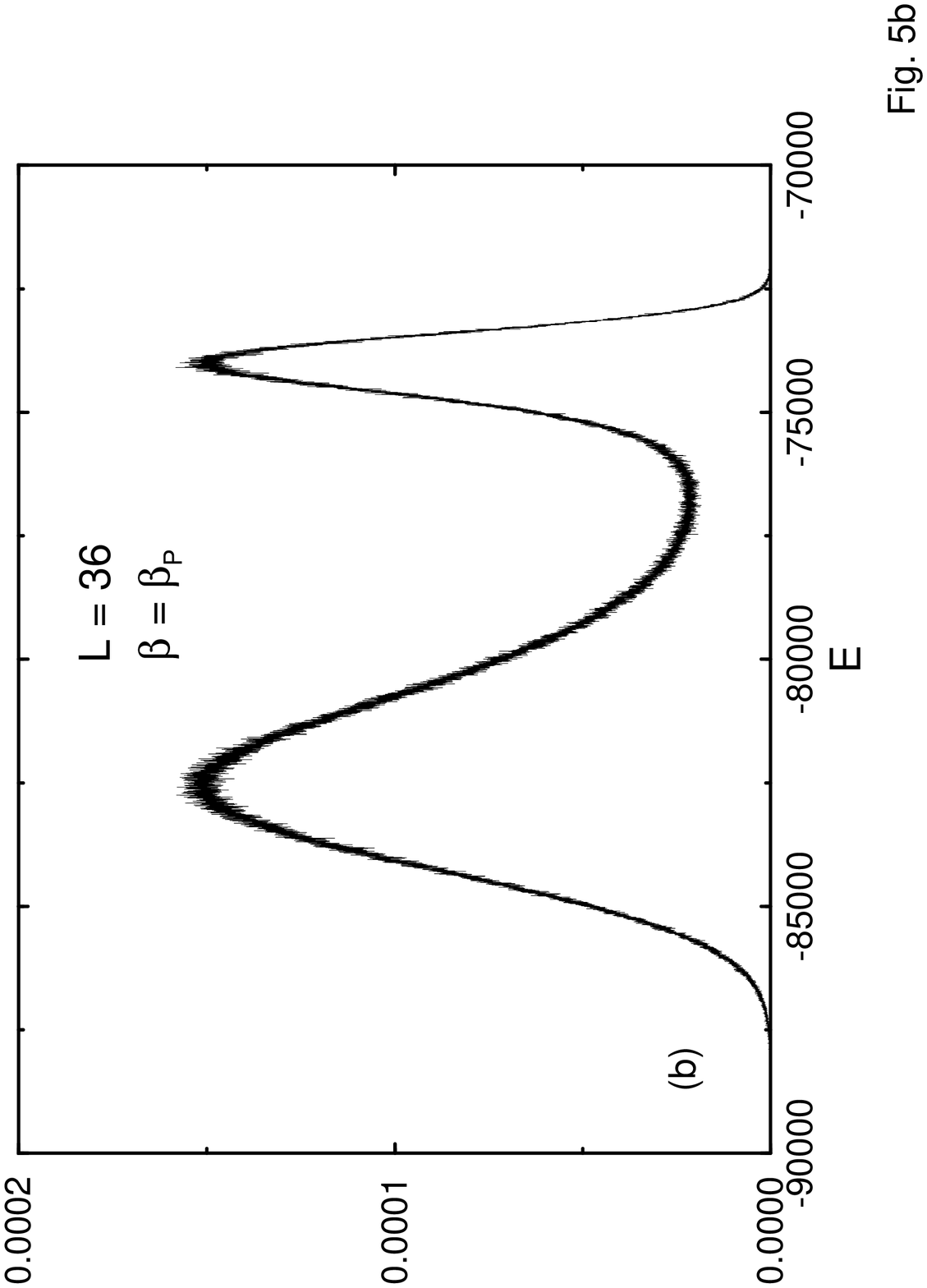}
\caption[a]{{\em Histogram for $L=36$ reweighted to
(a) $\beta_w(L=36) = 0.550\,581\,2$ and (b) $\beta_p(L=36) = 0.550\,566\,5$.}}
\label{fig:eqh_his}
\end{figure}

This completely unexpected result can be understood as follows:
The decomposition (\ref{eq:zper}) of the partition function translates
to the probability distribution (formally via an inverse Laplace 
transformation) at $\beta_t$ in a Gaussian approximation to
\begin{eqnarray}
P_{\beta_t}(e) &=& q \sqrt{\frac{V\beta_t^2}{2\pi c_o}}
                     \exp\left[-\frac{(e-e_o)^2 \beta_t V}{2 c_o}\right]
                +    \sqrt{\frac{V\beta_t^2}{2\pi c_d}}
                     \exp\left[-\frac{(e-e_d)^2 \beta_t V}{2 c_d}\right] 
                     \nonumber \\
               & \equiv & H_o(e) + H_d(e),
\label{eq:P}
\end{eqnarray}
where $e_o$ ($e_d$) and $c_o$ ($c_d$) are the energy and specific heat in
the ordered (disordered) phase at $\beta_t$, and $H_o$ ($H_d$) approximates
the ``ordered'' (``disordered'') part of the measured histograms.%
\footnote{In practice the separation of a given histogram into $H_o$ and
$H_d$ is of course not unique. Using as dividing energy 
the cut $E_0$ introduced earlier, then $\sum_e H_o(e) = W_o$ and
$\sum_e H_d(e) = W_d$. The Gaussian approximation is therefore a good
approximation in the sense that $W_o/W_d = q$ at $\beta_t$.} 
This implies for the ratio of heights at $\beta_t$
\begin{equation}
 R_H(\beta_t) \equiv \frac{H_o^{\rm max}}{H_d^{\rm max}} 
                   = q \sqrt{\frac{c_d}{c_o}}.
\label{eq:R_H}
\end{equation}
Estimates of $c_o$ and $c_d$ have recently been obtained in Ref.~\cite{jk96}
from simulations of a $100^3$ lattice in the pure phases at 
$\beta = 0.550\,53$, the previously best estimate of $\beta_t$. If we
perform a slight reweighting to our present best estimate of 
$\beta_t = 0.550\,565$ (see the Conclusions below) and insert the 
resulting numbers ($c_o = 28.84$, $c_d=3.210$) into (\ref{eq:R_H}) we obtain
\begin{equation}
R_H(\beta_t) = 1.001(18),
\label{eq:R_Hvalue}
\end{equation}
where the error emerges from the statistical uncertainties in $c_o$ and
$c_d$. The error due to the uncertainty in $\beta_t$ is of the same order;
see Fig.~\ref{fig:height_ratio}, where $R_H(\beta)$ is plotted over a range of
$\beta$ values around $\beta_t$. We thus have the surprising result that
for the three-dimensional 3-state Potts model the first-order transition
point $\beta_t$ is not only characterized by the, on quite general grounds, 
theoretically predicted ratio-of-weights $R_W(\beta_t) = q = 3$, but 
empirically also by an equal-peak-height condition, $R_H(\beta_t) = 1$.
This is qualitatively the reason why $\beta_P$ happens to be so close to
the infinite-volume transition point $\beta_t$. The actually measured 
probability distributions for $L=36$ reweighted to 
(a) $\beta = \beta_w(L=36) = 0.550\,581\,2$ and 
(b) $\beta = \beta_p(L=36) = 0.550\,566\,5$ are shown in 
Fig.~\ref{fig:eqh_his}. Notice the sensitivity to small variations in 
$\beta$ of the order of 2 -- 3 error bars of $\beta_w$ or $\beta_p$.

The result $c_o/c_d \approx 9 = q^2$ is presumably an accidental peculiarity
of this particular 3D 3-state Potts model. For 2D $q$-state Potts models with
$q \ge 5$ it is exactly known that $c_o \approx c_d$ at $\beta_t$, such that 
$R_H(\beta_t) \approx q$, and also for the 3D $q$-state models with $q=4$ 
and 5 one finds values of $R_H(\beta_t) \approx 1.4$ ($q=4$) and 
$R_H(\beta_t) \approx 1.8$ ($q=5$) (using again the estimates for $c_o$ and 
$c_d$ of Ref.~\cite{jk96}) which clearly deviate from unity. On the other 
hand, in (\ref{eq:R_Hvalue}) unity is hit so accurately that it is tempting
to speculate that there could be some hidden theoretical reason 
for this numerical observation (possibly related to the $Z_3$ symmetry 
of the 3-state Potts model). It would be very interesting to test this
possibility by studying the 3-state model on other three-dimensional lattices
(BCC, FCC, \dots) or in higher dimensions (4D, 5D, \dots). 

\subsection{Locations of $C_{\rm max}$}
\begin{figure}[bhp]
\vskip 7.0truecm
\includegraphics{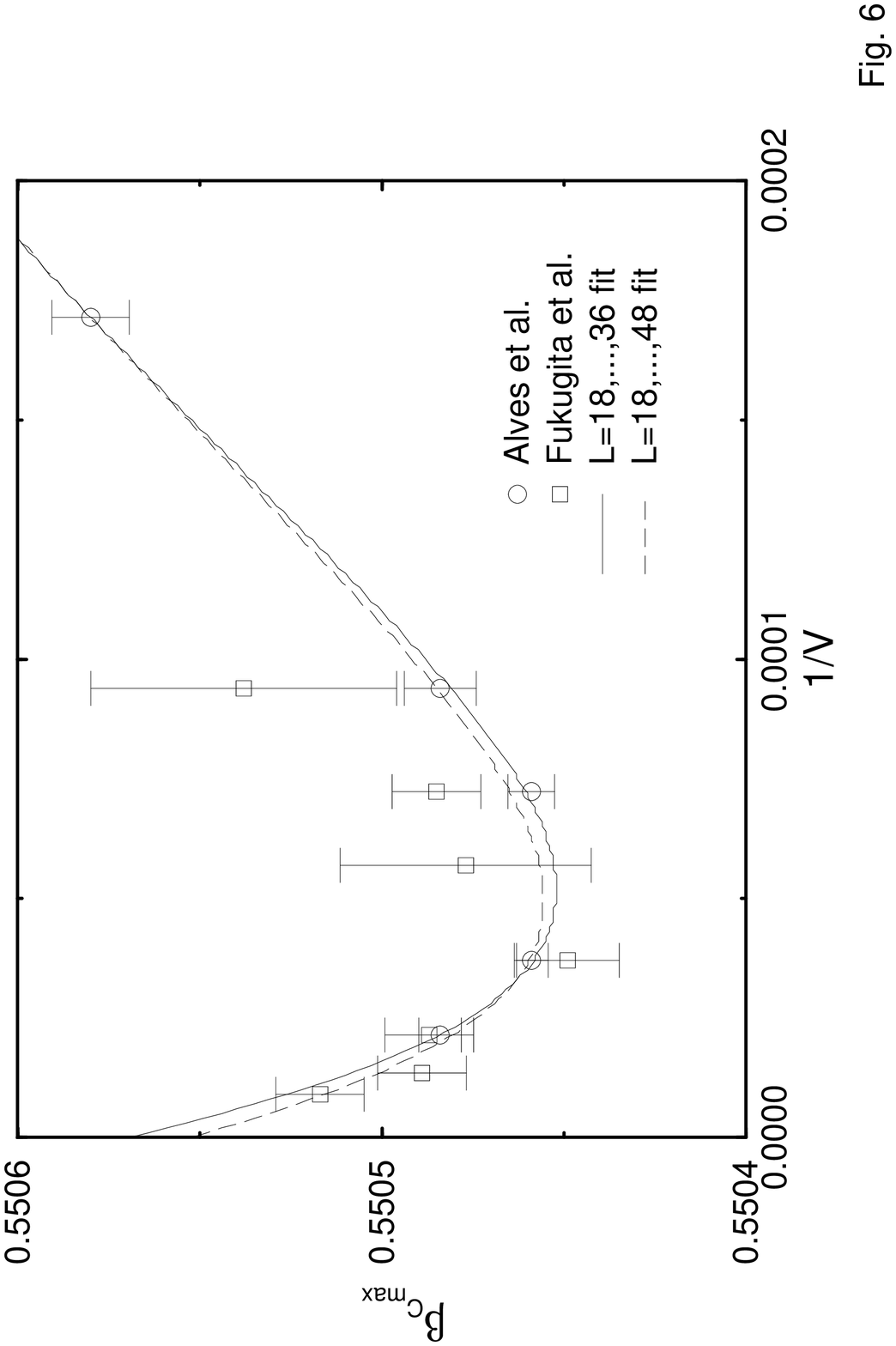}
\caption[a]{{\em Finite-size scaling of the specific-heat maxima
locations together with fits of the form $\beta_{C_{\rm max}} = \beta_t
+c/V + a \exp(-b L)$.}}
\label{fig:b_cmax}
\end{figure}

Next we also reconsider the scaling of the locations 
$\beta_{C_{\rm max}}$ of the specific-heat maxima, showing the peculiar
dip around $L=30$. As discussed in the Introduction a possible explanation
is a crossover between exponential corrections $\propto \exp(-L/L_0)$ 
and the asymptotic power series in $1/V$. We therefore tried to fit the
data with an ansatz
\begin{equation}
\beta_{C_{\rm max}}(L) = \beta_t + \frac{c}{V} + a_3 e^{-b_3 L}.
\label{eq:fit3}
\end{equation}
By following the procedure of ABV and
including also the data of Ref.~\cite{fmou90}, but discarding
only the three points with $L \le 14$, we obtained
\begin{equation}
  \beta_t = 0.550\,552 \pm 0.000\,020
  \mbox{~~~~($C_{\rm max}$ locations for $L \le 48$)},
\label{eq:bcmax_2}
\end{equation}
$c = -4.1(1.3)$, $a_3 = 0.0309(34)$, and
$b_3 = 0.207(16)$, with $Q=0.46$. 
The fit is shown as the dashed line on the fine scale of Fig.~\ref{fig:b_cmax}.
By including also the $L=14$ data in the fit, the numbers change only
slightly to $\beta_t = 0.550\,535(12)$, $c = -2.86(59)$, $a_3 = 0.0356(25)$,
and $b_3 = 0.2324(82)$, with $Q=0.41$.%
\footnote{It should be noted, however, that
the ansatz (\ref{eq:fit3}) admits
also another fit with ``unreasonable'' parameters ($c > 0$ and $a_3 < 0$)
which actually has even a slightly better $Q$ value of $Q=0.43$.}

One interesting question was if now already the data of ABV up to $L=36$ 
would be sufficient to get a reliable estimate of $\beta_t$. Using their five 
data points for $L=18,\dots,36$, we indeed obtained a reasonable fit with
\begin{equation}
  \beta_t = 0.550\,569 \pm 0.000\,042 
  \mbox{~~~~($C_{\rm max}$ locations for $L \le 36$)},
\label{eq:bcmax_1}
\end{equation}
$c = -4.9(2.6)$, $a_3 = 0.0351(53)$, $b_3 = 0.207(21)$, and $Q=0.71$. 
As can be inspected in
Fig.~\ref{fig:b_cmax} the two fits are in fact hardly distinguishable.

As a consequence of the additional exponential term in the fit ansatz
(\ref{eq:fit3}) the estimates of $\beta_t$ in (\ref{eq:bcmax_2}) and 
(\ref{eq:bcmax_1}) are somewhat higher than the result of ABV obtained 
from a pure power-law ansatz. In particular the estimate (\ref{eq:bcmax_1})
relying {\em only\/} on the data of ABV is now in perfect agreement with 
the results from the new methods discussed above. It appears quite natural
that the mixing with the data of Ref.~\cite{fmou90} perturbs this 
self-consistency a little bit.

Even though the high values of $Q$ indicate that both fits are 
statistically self-consistent, it is a priori clear that (\ref{eq:fit3})
can only be an effective model of the true behaviour, since higher order
terms in $1/V$ are neglected. A simple way to test for this is to compare the
coefficient $c$ of the $1/V$ term with the prediction of FSS theory 
\cite{jan93,kos}, $c = -\ln q/\Delta e \approx -6.5$, where $\Delta e$ can 
be taken from ABV or from the analysis in the next section.
We see that within the relatively large error bars the fitted values of $c$ are 
compatible with this expectation, but there is definitely a trend to
underestimate $c$. We have checked that the pure power-law fit used by ABV, 
$\beta_{C_{\rm max}} = \beta_t + c/V$, has the same problem; we
can reproduce their $\beta_t$ to all digits and obtain $c = -1.77(33)$. This
shows that the additional exponential correction helps at least to drive
$c$ in the right direction. A glance at Fig.~\ref{fig:b_cmax} then shows
that a pure power-law fit must necessarily underestimate $\beta_t$. 

\begin{figure}[bhp]
\vskip 7.0truecm
\includegraphics{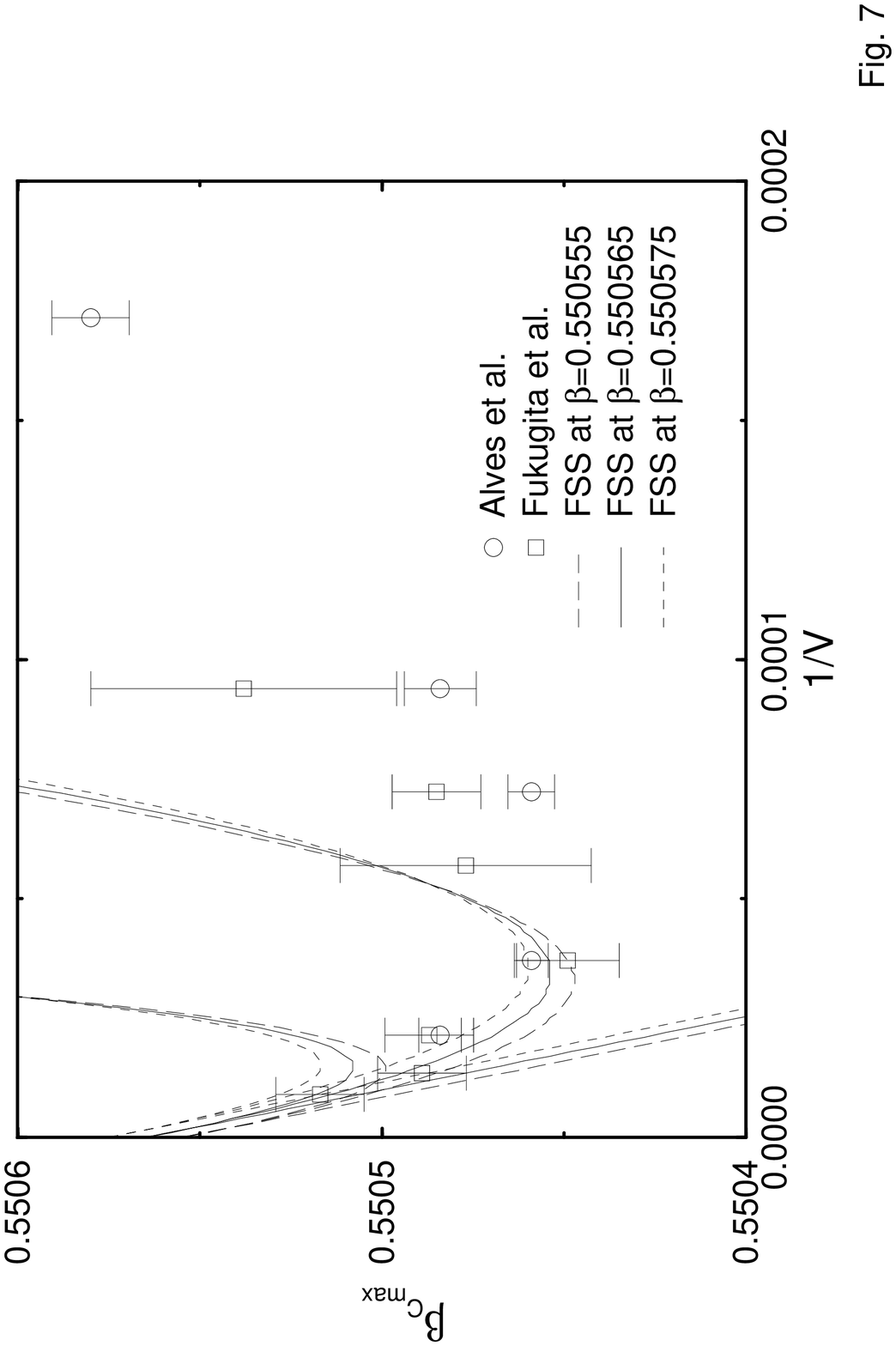}
\caption[a]{{\em Finite-size scaling of the specific-heat maxima
locations together with the theoretically predicted
FSS power-law expansion in $1/V$ to first, second, and third order (see text).}}
\label{fig:b_cmax_fss}
\end{figure}

How about higher-order corrections in the $1/V$ expansion?
The FSS expansion in $1/V$ up to second order reads \cite{jan93,kos}
\begin{equation}
\beta_{C_{\rm max}} = \beta_t - \frac{\beta_t \ln q}{V \Delta s}
                    + \frac{\beta_t}{(V \Delta s)^2}
                      \left[4 - 6 \frac{\Delta c}{\Delta s} 
                              +   \frac{\Delta c}{2 \Delta s} (\ln q)^2 \right] 
                    + {\cal O}(1/V^3),
\label{eq:fss}
\end{equation}
where $\Delta s = \beta_t \Delta e$ is the entropy jump, $\Delta e = e_d - e_o$
the latent heat, $\Delta c = c_d - c_o$ the specific-heat jump, and
$e_o$, $c_o$ etc. are to be taken as the (pure phase) expectation
values at $\beta_t$. Inserting $\beta_t = 0.550\,565$ and again the estimates
of Ref.~\cite{jk96}, we obtain the explicit FSS prediction
\begin{equation}
\beta_{C_{\rm max}} = 0.550\,565 - 6.502\,55/V + 94\,804.0/V^2 + \dots. 
\label{eq:fss_theory}
\end{equation}
In Fig.~\ref{fig:b_cmax_fss} the resulting first- and second-order FSS 
scaling curves are displayed as the solid lines on the same scale as in 
Fig.~\ref{fig:b_cmax}. As a rough error estimate the dashed lines show 
Eq.~(\ref{eq:fss}) evaluated at the error bounds of $\beta_t$, i.e., at
$0.550\,565 \pm 0.000\,010$. The statistical error coming from the
uncertainties in the pure phase expectation values is of the same order.
At first sight one is inclined to conclude that the second-order FSS curve
tends to reproduce the dip in the data. At its minimum, however, the
difference between the FSS expansions to first and second order is already
quite large, and one may suspect that the omitted terms of ${\cal O}(1/V^3)$ 
would change the picture drastically. In fact, by extending the expansion of
$\beta_{C_{\rm max}}$ up to the fourth order in $1/V$ \cite{jan_unp}
we clearly observe the breakdown of this asymptotic expansion for
moderate values of $V$. A similar observation was recently made in 
Ref.~\cite{morel} for the maxima locations of  
$C/\beta^2 = V (\langle e^2 \rangle - \langle e \rangle^2)$ 
in the two-dimensional 10-state Potts model, using large-$q$ expansions to
estimate the various energy cumulants involved.
%
%
\subsection{Latent heat and interface tension}
\begin{figure}[bhp]
\vskip 7.0truecm
\includegraphics{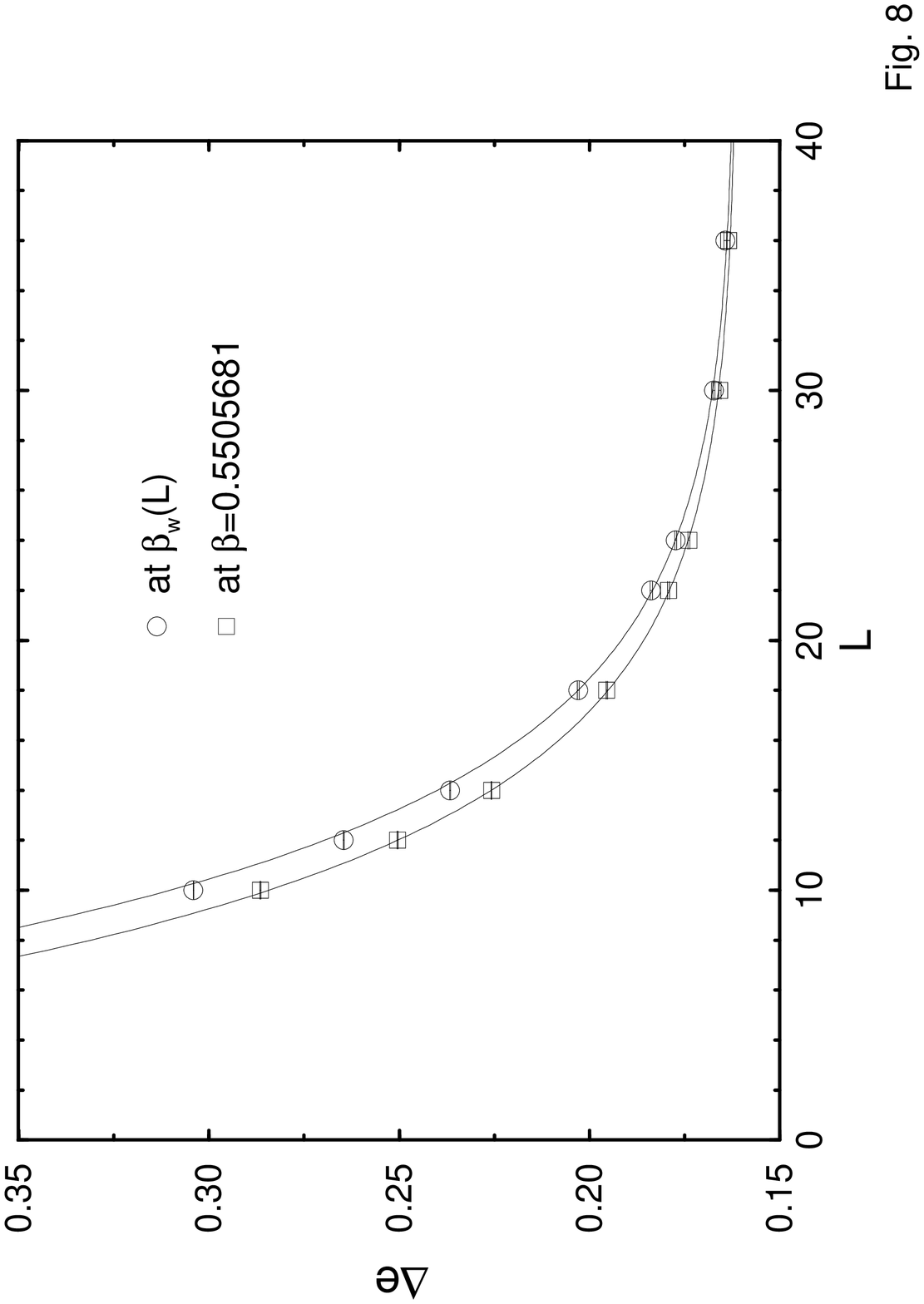}
\caption[a]{{\em Finite-size scaling of the latent heat derived from
the ratio-of-weights method. The continuous lines are fits of the form
$\Delta e(L) = \Delta e(\infty) + a \exp(-b L)$.}}
\label{fig:de_from_w}
\end{figure}

As proposed in Ref.~\cite{jan93} the ratio-of-weights method provides 
at $\beta_t$ also a natural estimator for the latent heat, 
\begin{equation}
  \Delta e(L) = \frac{d}{d\beta} \ln (W_o/W_d)/V,
  \label{eq:de1}
\end{equation}
which, similar to $\beta_w(L)$, should show only exponential deviations from
the infinite-volume limit, $\Delta e(\infty) = e_d - e_o$.
Graphically, the $V \Delta e(L)$ are just the slopes of $\ln R_W(\beta)
= \ln (W_o/W_d)$ shown in Fig.~\ref{fig:w_ratio} which can, of course, be
computed in precisely this way by numerical differentiation. It is, however,
more convenient and numerically more stable to perform the differentiation
in (\ref{eq:de1}) explicitly. Recalling the definition (\ref{eq:R_W}) 
it is easy to see that
\begin{equation}
  \Delta e(L) = \langle e \rangle_d - \langle e \rangle_o,
  \label{eq:de2}
\end{equation}
where $\langle e \rangle_d \equiv \frac{1}{V} \sum_{E \ge E_0} E P_\beta(E) /
\sum_{E \ge E_0} P_\beta(E)$ is the expectation value of the energy
computed over the disordered peak only, and $\langle e \rangle_o$ is defined
analogously. We computed $\Delta e(L)$ in two different ways. 
First, at the infinite-volume limit of $\beta_w(L)$,
$\beta = 0.550\,568\,1 \approx \beta_t$, and second self-consistently at
the lattice size dependent sequence of $\beta = \beta_w(L)$. The resulting
data is given in Table~\ref{tab:delta_e} and shown in Fig.~\ref{fig:de_from_w}.
%
%
%
\begin{table}[tb]
\centering
\caption[{\em Nothing.}]
 {{\em Finite lattice estimates for the latent heat using the
   ratio-of-weights method (with $\beta_w(\infty) = 0.550\,568\,1$)}}
\vspace{3ex}
\begin{tabular}{r|ll}
\multicolumn{1}{c|}{$L$}  &
\multicolumn{1}{c}{$\Delta e$ at $\beta_w(\infty)$}  &
\multicolumn{1}{c}{$\Delta e$ at $\beta_w(L)$} \\
\hline
 10 &  0.286\,50(12)  &    0.304\,04(12) \\
 12 &  0.250\,45(13)  &    0.264\,62(14) \\
 $14$ &  0.225\,76(10)  &  0.236\,66(10) \\
 18 &  0.195\,49(23)  &    0.202\,97(23) \\
 22 &  0.179\,26(28)  &    0.183\,80(28) \\
 24 &  0.173\,92(23)  &    0.177\,39(23) \\
 30 &  0.165\,80(30)  &    0.167\,34(30) \\
 36 &  0.163\,43(37)  &    0.164\,33(36) \\
\end{tabular}
\label{tab:delta_e}
\end{table}

The fits in Fig.~\ref{fig:de_from_w} are according to the ansatz
\begin{equation}
  \Delta e(L) =  \Delta e + a_4 e^{-b_4 L},
  \label{eq:fit4}
\end{equation}
where $\Delta e \equiv \Delta e(\infty)$ is the infinite-volume limit of the
latent heat. When $\beta =0.550\,568\,1 \approx \beta_t$ is held fixed,
$\Delta e(L)$ follows the exponential ansatz down to quite small lattice 
sizes, and by discarding only the $L=10$ data point we obtain the 
infinite-volume estimate
\begin{equation}
  \Delta e = 0.161\,21 \pm 0.000\,27,
  \label{eq:de_at_bt}
\end{equation}
$a_4 = 0.6160(80)$, and $b_4 = 0.1611(12)$, with $Q=0.14$. When the
$\Delta e(L)$ are evaluated at $\beta_w(L)$, the asymptotic scaling region
is shifted to larger values of $L$. In order to get a fit of reasonable 
quality, here we had to discard the data for $L \le 14$. The result of the fit, 
\begin{equation}
  \Delta e = 0.161\,60 \pm 0.000\,47,
  \label{eq:de_at_bw}
\end{equation}
with $a_4 = 0.733(50)$, $b_4 = 0.1597(42)$, and $Q=0.11$, is then
compatible with (\ref{eq:de_at_bt}), and both estimates are consistent
with the previous estimate of ABV from the scaling of the specific-heat
maxima. 

Let us finally consider the (reduced) interface tension $\sigma_{od}$. As can
be read off from Table~\ref{tab:beta_w}, the data for $L=12,\dots,36$ vary so
little that none of the more sophisticated FSS extrapolations is applicable.
Similar to the analysis of $\beta_p$ we therefore have again simply taken
a weighted average of the seven values with the result
\begin{equation}
2 \sigma_{od} = 0.001\,63 \pm 0.000\,02,
\label{eq:sigma_val}
\end{equation}
and $Q=0.87$. A similar estimate of $2 \sigma_{od} = 0.001\,57(6)$ was 
recently obtained in Ref.~\cite{schmidt}.
%
\section{Conclusions}
%
In this paper we have presented a detailed study of the first-order phase
transition in the three-dimensional 3-state Potts model on cubic lattices
with emphasis on recently introduced quantities whose infinite-volume 
extrapolations are governed {\em only} by exponentially small terms. The 
main results are the following:

(i) The phase transition is only weakly first-order. Nevertheless the new
techniques, originally introduced for ``large enough'' $q$, i.e., strong
first-order phase transitions, prove to be surprisingly accurate. The
expected exponential finite-size scaling of the pseudo-transition points
$\beta_{V/V'}$ and $\beta_w$ as well as of the ``ratio-of-weights'' definition
of the latent heat is clearly observed. The results of fits of the form
$\beta_t(L) = \beta_t + a\exp(-b L)$ to the various sequences of
pseudo-transition points are summarized in Table~\ref{tab:betas}.
Taking into account the small remaining systematic uncertainties by 
averaging over the entries in Table~\ref{tab:betas} we quote as
our final estimate for the infinite-volume transition point
\begin{equation}
\beta_t = 0.550\,565 \pm 0.000\,010.
\end{equation}

(ii) The points $\beta_p$, where the two peaks of the energy probability
distribution are of equal height, show surprisingly almost no finite-size
scaling. This in turn implies that in this particular model the transition 
point is not only characterized by the theoretically founded 
``3:1-weight rule'', but empirically to a very good approximation also 
by an ``equal-height rule''. We emphasize that this could be a purely 
accidental situation since an ``equal-height rule'' is definitely not 
satisfied in two-dimensional $q$-state Potts models for all $q \ge 5$ and 
in the three-dimensional models with $q=4$ and $5$.

%
%
\begin{table}[tb]
\centering
\caption[{\em Nothing.}]
 {{\em  Summary of our various estimates of $\beta_t$ }}
\vspace{3ex}
\begin{tabular}{|l|l|l|l|l|}
\hline
\multicolumn{1}{|c|}{obs.}        &
\multicolumn{1}{c|}{range}       &
\multicolumn{1}{c|}{ansatz}      &
\multicolumn{1}{c|}{$\beta_t$}   &
\multicolumn{1}{c|}{$Q$}         \\
\hline
$\beta_{V/V'}$ & $L'=36$, $L=12-30$  & $\beta_t + a e^{-bL}$
                                        &   0.550\,586(10)  & 0.99 \\[0.4cm]
$\beta_{V/V'}$ & $L'=30$, $L=12-24$  & $\beta_t + a e^{-bL}$
                                        &   0.550\,590(20)  & 0.95 \\[0.4cm]
$\beta_{V/V'}$ & diagonal                  & $\beta_t + a e^{-bL}$
                                        &   0.550\,586(15)  & 0.79 \\[0.2cm]
\hline
$\beta_w$ & $L=14-36$                & $\beta_t + a e^{-bL}$
                                        &   0.550\,568\,1(56) & 0.41 \\[0.2cm]
\hline
$\beta_P$ & $L=14-36$                &  average
                                        &   0.550\,562\,5(36) & 0.16  \\[0.2cm]
\hline
$\beta_{C_{\rm max}}$ & $L=14-48$    & $\beta_t + c/V + a e^{-bL}$
                                        &   0.550\,535(12)    & 0.41 \\[0.4cm]
$\beta_{C_{\rm max}}$ & $L=18-48$    & $\beta_t + c/V + a e^{-bL}$
                                        &   0.550\,552(20)    & 0.46 \\[0.4cm]
$\beta_{C_{\rm max}}$ & $L=18-36$    & $\beta_t + c/V + a e^{-bL}$
                                        &   0.550\,569(42)    & 0.71 \\[0.2cm]
\hline
\end{tabular}
\label{tab:betas}
\end{table}

(iii) The peculiar dip in the finite-size scaling of the locations of the
specific-heat maxima can be explained by a fit of the form
$\beta_{C_{\rm max}}(L) = c/V + a \exp(-b L)$.

(iv) The decay constant $b$ in the exponential term $\exp(-b L)$ is quite
consistently found from the various fits around $b \approx 0.2$. This implies
$L_0 = 1/b \approx 5$, which is about one-half of the correlation length 
$\xi_d \approx \xi_o \approx 10$ in the pure disordered and ordered phases.
A similar relation was observed in Refs.~\cite{blm1,icke-micro} for the
two-dimensional 10-state Potts model, where $\xi_d$ is known exactly and
evidence for $\xi_o = \xi_d$ was obtained numerically with high 
accuracy \cite{our_epl}.

(v) Similar to the discussion of the two-dimensional 10-state Potts model
in Ref.~\cite{morel}, we find that the asymptotic expansion in 
$1/V$ for $\beta_{C_{\rm max}}$ is ill-behaved. Only for very large system 
sizes higher order terms lead to an improvement which, however, is quite a 
general feature of asymptotic expansions. 

(vi) The finite-size scaling of the ``ratio-of-weights'' definition of the 
latent heat follows the predicted exponential behaviour. The resulting 
infinite-volume estimate,
\begin{equation}
\Delta e = 0.161\,4 \pm 0.000\,3,
\end{equation}
was found compatible with the previous estimate of ABV from the finite-size
scaling of the specific-heat maxima.

(vii) The interface tension $\sigma_{od}$ between the ordered and disordered
phases has probably not yet reached its asymptotic scaling region for 
lattice sizes up to $L=36$. Experience with three-dimensional $q$-state
Potts models with $q=4$ and $5$ suggests \cite{malcolm} that our estimate 
$2 \sigma_{od} = 0.001\,63(2)$ is a lower bound on the infinite-volume limit.
Still it is difficult to reconcile this value with the estimates of
$\sigma_{oo}(\beta_t) \ge 0.0048(3)$ and $\sigma_{oo}(\beta_t) = 0.0040(7)$ 
in Ref.~\cite{patkos} and the quite general stability condition 
$\sigma_{oo}(\beta_t) \le 2 \sigma_{od}$ \cite{scho}. Notice that at least 
in two dimensions equality seems to hold in the last relation (``complete 
wetting''). 
%
\section*{Acknowledgements}
%
We would like to thank Stefan Kappler for many useful discussions.
This work was supported in part by a collaborative research grant within the 
Programme ``Acciones Integradas'' of the DAAD/Spanish Ministry of Education.
WJ thanks the Deutsche Forschungsgemeinschaft (DFG) for a Heisenberg 
fellowship, and RV acknowledges partial support by CICYT under contract 
AEN95-0882.
%
%
\newpage
%

%
%
\newpage
\clearpage\newpage
\begin{figure}[bhp]
\vskip 7.0truecm
\includegraphics{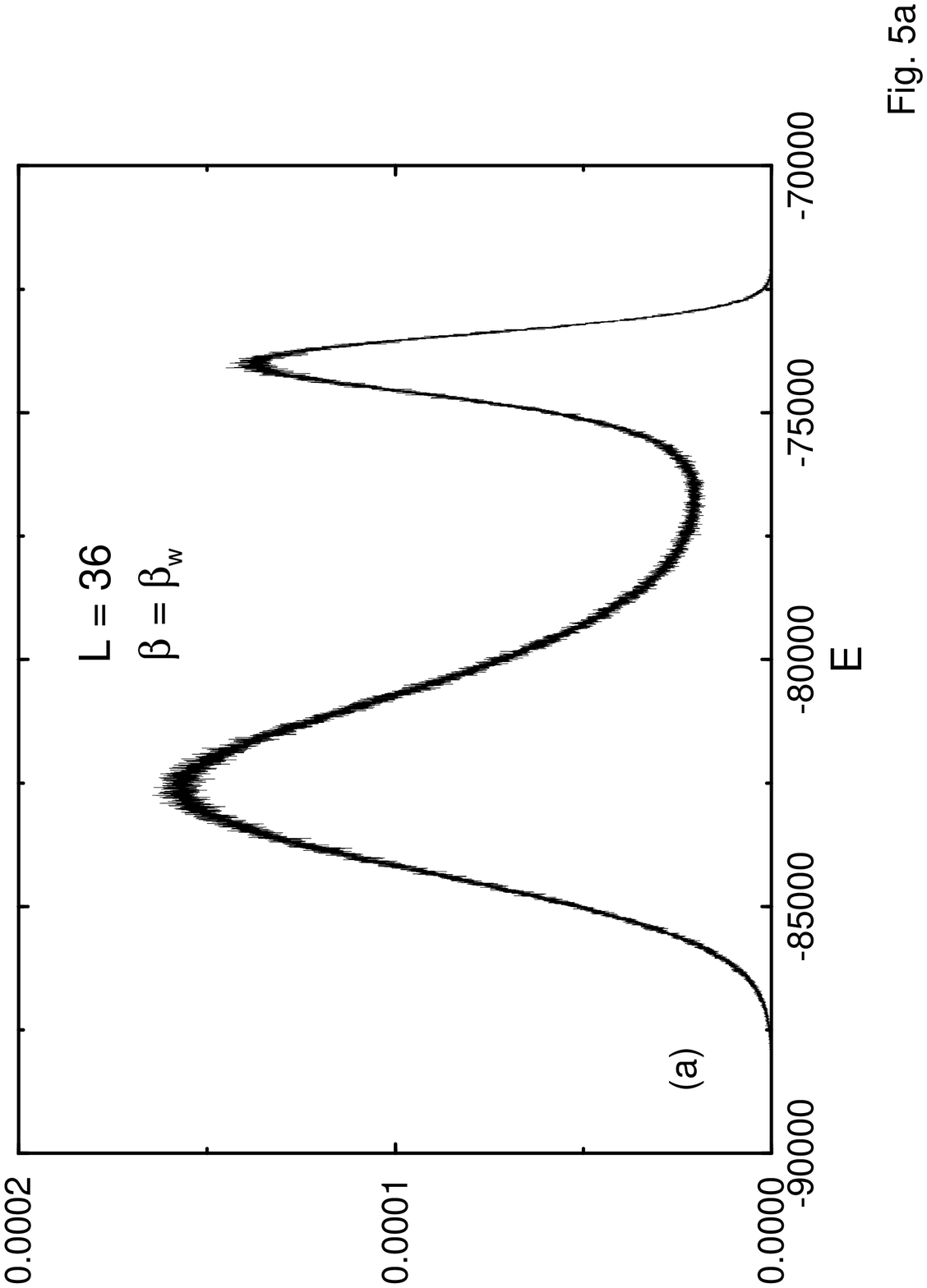}
\label{fig:eqw_his}
\end{figure}
\end{document}